# Influencing Optical and Charge Transport Properties by Controlling the Molecular Interactions of Merocyanine Thin Films

*Lukas Böhner[1], Philipp Weitkamp[1], Thorsten Limböck[1], Nora Gildemeister[1], Daniele Fazzi[1,2], Manuela Schiek[3], Ruth Bruker[1], Dirk Hertel[1], Roland Schäfer[1], Klas Lindfors[1], and Klaus Meerholz[1]\**

[1]Department of Chemistry, University of Cologne
Greinstraße 4-6, 50939 Cologne, Germany

[2] Universita di Bologna, Dipartimento di Chimica "Giacomo Ciamician",
Via P. Gobetti 85, 40129 Bologna, Italy

[3] ZONA
Johannes Kepler Universität Linz,
Linz 4040, Austria

\* klaus.meerholz@uni-koeln.de

# Abstract

In organic semiconductors charge transport typically takes place via slow hopping processes. Molecular aggregation can lead to enhanced exciton and charge transport through coupling of the transition dipole moments. In this work, we investigate the optical, morphological, and electronic properties of thin films of a merocyanine dye, that aggregates due to its high ground-state dipole moment. The degree of aggregation of spin-coated thin films can be easily tuned by thermal annealing. We demonstrate the relationship between charge carrier mobility and the degree of aggregation. The mobility is increased by approximately three orders of magnitude due to aggregation. We combine variable angle spectroscopic ellipsometry and polarization-resolved absorption spectroscopy with density functional theory to demonstrate that the aggregated molecules are oriented in an upright, standing configuration relative to the substrate surface. This arrangement involves a co-facial orientation of the molecular pi-systems which is advantageous for lateral charge transport. By utilizing highly oriented pyrolytic graphite as an ordered substrate, we are able to template the growth of the merocyanine layer in vapor phase deposition, and to improve the in-plane morphological order drastically. By correlating atomic force microscopy and photoluminescence microspectroscopy we observe oriented domains of 100s of $\mu m^2$ in size, emitting linearly polarized light, whereby maintaining the edge-on molecular arrangement. This promises a further significant enhancement of lateral charge carrier mobility.

# Introduction

Organic semiconducting materials as active thin films in electronic devices are usually present in a polycrystalline or amorphous state. They mostly exhibit lower charge carrier mobilities than single crystals, due to comparably low long-range order. The highest charge carrier mobility values for organic small molecules are usually obtained for single crystals of rod-, or disk-like molecules such as copper-phthalocyanine (1.0 cm²/Vs),[1] tetracene (2.4 cm²/Vs),[2] pentacene, or rubrene (40 cm²/Vs)[3,4]. Although highly dipolar molecules were long expected to show low charge carrier mobilities associated with increased energetic disorder, a mobility of 2.3 cm²/Vs was measured for a merocyanine single crystal,[5] demonstrating the concept of molecular dipole elimination upon centrosymmetric packing.[6–8] However, the single-crystal growth process is lengthy and less feasible for utilization in organic electronic devices compared to amorphous or polycrystalline organic thin films, fabricated by, e.g., spin coating or thermal evaporation, especially for large area devices. Tremendous efforts have been undertaken into introducing a higher degree of order in such organic thin films to improve the charge transport performance while maintaining a homogeneous and dense film quality. Increasing order in organic thin films can be achieved by, e.g., templating substrates[9–12], external stimuli such as mechanical rubbing[13,14], or different liquid-phase deposition techniques which influence molecular orientation.[15–17]

Additionally, it was shown that molecular aggregation of polar dyes like merocyanines or squaraines can lead to improved charge carrier mobility.[18–21] In these aggregates excitons are delocalized by coupling of the transition dipole moments (TDM) of the associated monomers.[22,23] This leads to drastic changes of their optical properties compared to the monomeric form, which allows for monitoring the aggregation state by absorption or photoluminescence spectroscopy. The two best known forms of molecular aggregates are J- and H-aggregates.[24] J-aggregates are known to show intense, red-shifted and narrow absorption peaks relative to the monomer. Further, photoluminescence (PL) with negligible Stokes shift can be observed. By contrast, H-aggregates usually show a broad, blue-shifted, and vibronically rich absorption spectrum, and the emission is strongly quenched. The differences in the spectra of J- and H-aggregates arise from the orientations of the TDMs of the aggregated molecules relative to each other. In J-aggregates the TDMs are oriented in a head-to-tail fashion, whereas in H-aggregates they are oriented in a sandwich-type packing arrangement, leading to the formation of antiparallel dimers.[22-25]

Merocyanine molecules possess large absorption coefficients which makes them interesting for applications such as solar cells or photodiodes.[18,27] Additionally, the large dipole moment of merocyanines makes them prone to aggregation. We exploit this to investigate a merocyanine molecule, namely HB238 [6], with high ground-state dipole moment of 13.1 D, which aggregates in a controlled way by tuning the experimental conditions. We use these aggregates as a model system to investigate the interdependence of morphological, optical, and charge transport properties. We fabricate and characterize spin-coated thin films on oxide surfaces, and we are able to show that the charge carrier mobility is significantly improved upon aggregation. By preparing thermally evaporated thin films on highly oriented pyrolytic graphite (HOPG) we enhance the structural order even further. We observe large, directionally oriented domains of 100s of $\mu m^2$ in size. Our observations reveal that molecular order can be induced by the choice of a suitable substrate, potentially improving charge carrier transport.

## Results & Discussion

The investigated merocyanine molecule with the trivial name HB238[6] (chemical structure shown in the inset in **Figure 1a)** is present in the monomeric (i.e., non-aggregated) form in acetone solution (c = $10^{-5}$ M). Calculation of the single molecule vibronic absorption spectrum and Huang-Rhys parameters, as carried out at the DFT/TDDFT level of theory (see Experimental techniques and methods), well resembles the experimental UV-Vis data, supporting that HB238 is present as non-aggregated monomeric species in solution (see Figure S1). A typical mirror-image of absorption and emission in the monomeric form is observed. When water is added, dramatic changes in the absorption and photoluminescence spectra are observed (see **Figure 1a** and **b**). We attribute these observations to the formation of HB238 aggregates resulting from the insolubility of HB238 in water. Upon aggregation, two distinct peaks arise in the absorption spectrum. One of the peaks is blue- [H-band: 2.47eV, FWHM = 126 meV (1014 cm$^{-1}$)] and the other one is red-shifted [J-band: 1.67 eV, FWHM = 49 meV (398 cm$^{-1}$)] compared to the monomer absorption feature. A very weak absorption band reminiscent of the monomer transition can still be seen around 1.9 eV in the aggregate spectrum. Additionally, we observe a narrow emission peak [1.65 eV, FWHM = 40 meV (322 cm$^{-1}$)] which is almost resonant to the red-shifted absorption peak of the J-band.

These findings correlate well with an aggregated state, in which the molecular TDMs are aligned to each other with an oblique angle, a model already proposed by Kasha *et al.* in 1965.[22] The lower and higher lying states, resulting from the sum and difference of the oblique-angled TDMs, respectively, reveal the resulting exciton band bottom and top, giving rise to *Davydov* splitting (DS).[36] Unlike pure J- or H-aggregates, both states gain oscillator strength. Emission, however, is only observed from the lowest state, i.e., the J-state, due to the fast internal non-radiative conversion. The observed DS is large (ca. 0.8 eV) in comparison with values reported for other other organic compounds[37–41], ranging between approximately 0.14 meV for pentacene[38] to 0.65 meV for squaraines[40], demonstrating an unusually strong interaction between the molecules in our case. This interpretation is supported by the fact that the H-transition displays a spectrally narrow absorption peak without significant vibronic features, which is only observed for very strong coupling, when there is little spectral overlap between the monomer and H-band absorption (compare **Figure 1a** and **b**, respectively).[26]

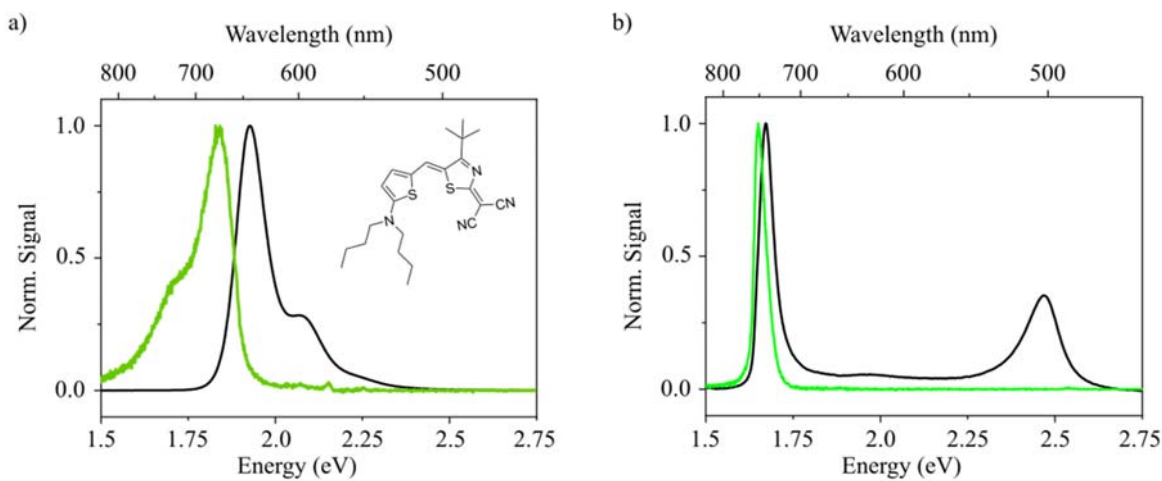

*Figure 1:* Optical properties of merocyanine HB238 in the monomeric and aggregated state. Normalized absorption and emission spectra (black and green lines, respectively) of monomeric HB238 in acetone solution (**a**) and dispersed aggregates in a 10 % acetone/90 % water mixture (**b**). The inset in panel **a** shows the chemical structure of HB238.

We now turn to spin-coated HB238 thin films necessary for device application. Spin coating on float glass substrates leads to thin films which show broad absorption spectra. The spectra can be significantly altered by thermally annealing the samples. To investigate in detail the annealing-dependent optical properties, spin-coated thin films were annealed for ten minutes at different temperatures in the range of 50 °C to 160 °C. An additional unannealed [referred to as "as fabricated (a.f.)"] sample was prepared. For every annealing temperature a separate sample was fabricated. The spin-coated thin films further show pronounced uniaxial anisotropy (direction different normal to the substrate, here z-direction, from substrate plane, here xy-plane) which is discussed further below in the main text in more detail. All spectra were thus measured at an angle of incidence (AOI) of 45° and with p-polarized light, to address the ordinary and extra-ordinary optical components with equal electric field strengths. The results are shown in **Figure 2**. The samples can be ordered into three regimes I, II, and III according to the annealing temperature. For clarity only a single absorption spectrum for each regime is shown in **Figure 2a** (all absorption spectra are shown in Figure S17).

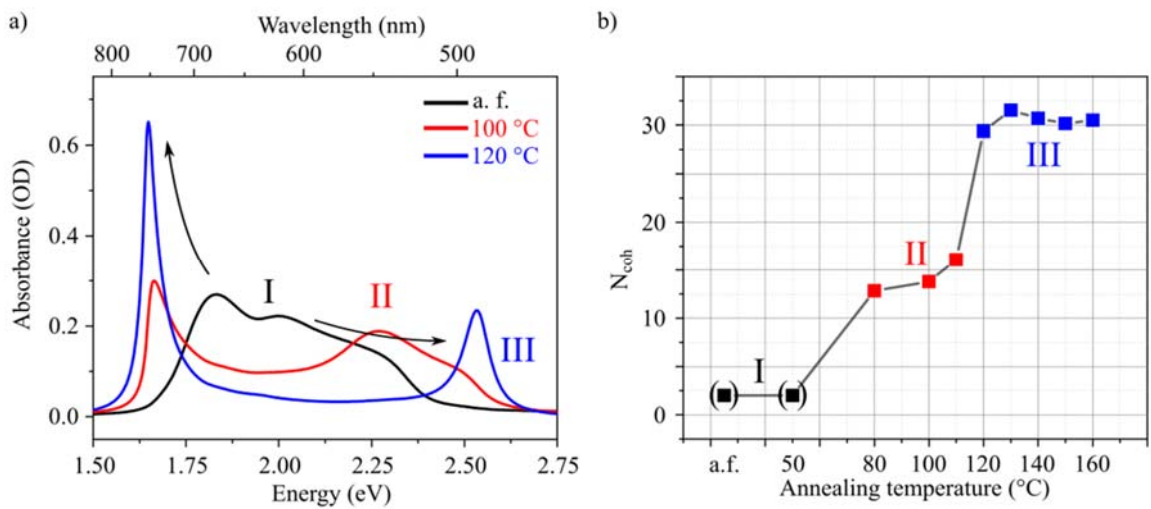

*Figure 2: Annealing-induced aggregation of HB238. (**a**) Absorption spectra illustrating the three different regimes (see main text) measured at 45° AOI using p-polarized light for the as fabricated sample (**I**, black), after annealing at 100 °C (**II**, red), and after annealing at 120 °C (**III**, blue). Black arrows are guides to the eye emphasizing spectral shifts. (**b**) Estimated number of coherently coupled molecules as a function of annealing temperature. The three different regimes can be distinguished and are highlighted with the same color as in (**a**).*

For **regime I** (low annealing temperatures; a.f. and 50 °C), a broad absorption band without clear peaks is observed. At intermediate temperatures (80 °C – 110 °C), **regime II**, the spectra display the characteristic J-transition peak at ca. 1.65 eV. Additionally, the H-transition is observed as a spectral shoulder at around 2.50 eV. However, the spectra also show broad features, mainly at around 2.25 eV, which overlap with the H-transition absorption peak. Annealing at 120 °C or higher (here: up to 160 °C) leads to **regime III**. The absorption spectra are dominated by distinct J- and H-transition peaks. The samples of **regime III** are thus in a similar aggregation state as the dispersed aggregates presented in **Figure 1b**. This is further supported by 2θ XRD measurements of both samples (see Figure S3). From **Figure 2a** it can be observed that, from **regime I** to **III**, the lowest energy vibronic peak at approximately 1.8 eV red shifts and finally results in the observed J-transition peak. Likewise, the higher energy vibronic transitions blue shift to finally evolve in the distinct H-transition peak. This evolution is indicated by the black arrows in **Figure 2a**. A similar behavior has been described by *Spano,* and associated to increasing coupling strength[23]. We thus attribute the observed changes in the absorption spectra during the transitions between the regimes to increased intermolecular coupling. The spectral reorganization and thus coupling strength do not change gradually with annealing temperature, but stepwise between the three regimes. We attribute this finding to thermodynamic boundaries which are overcome, leading to molecular reorganization between 50 °C and 80 °C (**I** to **II**) and between 110 °C and 120 °C (**II** to **III**). This is further supported by 2θ XRD measurements of the thin films shown in Figure S4. In **regime I** there is no measurable diffraction peak, in **regime II** a diffraction peak of low intensity is observed, and from **regime II** to **regime III** the intensity of the peak suddenly increases by an order of magnitude. The peak intensity saturates in **regime III** in accordance with the unaltered spectral properties **in regime III** (see Figure S17).

To further quantify the degree of aggregation, we estimate the number of coherently coupled molecules ($N_{coh}$) of each sample using the absorption spectra. The parameter $N_{coh}$ describes the number of molecules over which, on average, the exciton is delocalized, and is thus not typically equal to aggregate, or crystal domain dimensions. To estimate this number from absorption spectra, we utilize a method introduced by *Knoester* and *Bakalis*,[28] which was developed for linear J-aggregates. The method assumes that the spectral width of the J-peak [as half-width at half-maximum (HWHM)] corresponds to the energetic difference between the lowest (J-state) and the second lowest state of the exciton-band. With this assumption, the peak-width can be directly related to the number of coherently coupled molecules. For HB238-aggregates the mathematical description of a linear J-aggregate's exciton band is insufficient. As shown in, e.g., **Figure 1**, HB238-aggregates display two distinct peaks in the absorption spectrum, split asymmetrically around the monomer-absorption, with the H-transition absorption peak shifted further. These spectral features can be described using the exciton band description of two-dimensional, square herringbone aggregates developed by *Spano*.[42]

Using the model of Ref. 42, the number of coherently coupled molecules $N_{coh}$ can be related to the spectral width of the J- and H-transition and expressed as

$$N_{coh} = \left(\frac{\pi}{\cos^{-1}\left(\frac{HWHM_J}{-2(J_0-2J_1)}+1\right)}\right)^2, \qquad (\text{eq. 3})$$

for the J-transition and as

$$N_{coh} = \left(\frac{\pi}{\cos^{-1}\left(\frac{HWHM_H}{-2(J_0+2J_1)}+1\right)}\right)^2, \qquad (\text{eq. 4})$$

for the H-transition. Here $J_0$ and $J_1$ are the nearest and the next-nearest neighbor coupling constants, respectively, and $HWHM_{J(H)}$ is the half-width at half maximum of the J(H)-transition absorption peak. The HWHM values correspond to the splitting between the two lowest (J-transition; **Eq. 3**), or the two highest (H-transition; **Eq. 4**) exciton-states. This assumption is only valid, if the oscillator strength is mainly located in the exciton-band bottom and top, respectively. *Eisfeld* and *Briggs* showed, that this is the situation if the spectral overlap between the aggregate- and monomer-absorption peaks is negligible.[26] This condition, which is unusual for H-transition absorption peaks, is fulfilled by HB238 aggregates as shown in **Figure 1**. The HWHM values are directly obtained from the measured spectra (see Experimental techniques and methods for details of the peak fitting procedure, and Figure S21 for visualization of the fits). The nearest neighbor coupling $J_0$ is as well directly obtained from the aggregate absorption spectra from the davydov splitting.[42] The next-nearest neighbor coupling $J_1$ is obtained using the equality of **Eqs. 3** and **4**.

The advantage of this procedure is that all parameters are obtained from the same spectrum. It is not necessary to compare aggregate to monomer spectra which always includes uncertainties, due to the immeasurable, at least directly, liquid-to-crystal shift. The procedure yields an average number of coherently coupled molecules in a two-dimensional square aggregate. It has to be noted, however, that the described procedure includes some uncertainties. The HB238 aggregates might not be perfectly square. The AOI has a slight influence on the absorption peak-widths. Here all measurements were performed at an AOI of 45° with p-polarized light. The fitting of J- and H-transition absorption peaks of the samples annealed at 80 °C, 100 °C, and 110 °C is complicated by spectrally overlapping transitions. The obtained fits are still reasonable, evidenced by the fitted spectra shown in Figure S21. The estimated $N_{coh}$ of the spin-coated films are shown in **Figure 2b** as a function of annealing temperature. Due to the above discussed reasons these values should be taken as an effective, relative measure for the exciton delocalization, instead of absolute values.

With the described procedure, it is not possible to properly estimate $N_{coh}$ for the samples of **regime I** due to the absence of pronounced J- and H-aggregate peaks (see **Figure 2a**). We do however observe a slight anisotropic optical absorption behavior of these samples, which suggests that the thin films are not (entirely) composed of uncoupled monomers. Due to the high dipole moment of HB238 we assume the presence of a mixture of monomers and small molecular aggregates for the samples of **regime I**. For visualization, we set $N_{coh} = 2$ for the samples of **regime I**. In **regime II** the samples exhibit $N_{coh}$ values of around 15 slightly increasing with increasing temperature. Suddenly, in **regime III** at an annealing temperature of 120 °C, $N_{coh}$ drastically increases to a value of ca. 30 where it stays approximately constant for higher annealing temperatures. We additionally measured the XRD diffraction patterns of all samples (see Figure S4). We observe only one significant diffraction peak at $2\theta \approx 5.0°$, demonstrating that in all films in **regime II** and **III** a similar crystal structure is present and the dependence of its integrated intensity on the annealing temperature is in good agreement with the obtained $N_{coh}$. In **regime I** no significant diffraction peaks were measured, suggesting amorphous films. The samples in **regime II** mark the start of crystallization, slowly progressing with increasing annealing temperature within **regime II**, evidenced by the appearance of slowly growing diffraction peaks. Finally, the integrated diffraction peak intensity increases drastically during the transition from **regime II** to **III**, as a threshold annealing temperature for molecular reorganization is reached somewhere between 110 °C and 120 °C.

We conclude, that an $N_{coh}$ of ca. 30 represents a saturation value for spin-coated thin films of HB238, determined by an average crystal defect density and disorder. In comparison, $N_{coh}$ was calculated with the above-described procedure to be 22 for the dispersed aggregates shown in **Figure 1**. The lower $N_{coh}$ for dispersed aggregates might be explained with the instantaneous (kinetic) formation upon addition of water and an accompanied probability for packing defects.

We also tried annealing at higher temperatures. Starting from an annealing temperature of 170 °C, we observe pronounced depletion zones in the thin film morphology (dewetting), which become larger and more frequent at higher annealing temperatures (see Figures S8c-f and S9a-b). Concomitantly, the corresponding absorption spectra show a noticeable high-energy spectral shoulder of the J-transition absorption peak (see Figures S15 and S16a), which complicates the analysis of $N_{coh}$. Finally, at 200 °C the material melted, leaving only mono- and few-layer islands on the surface (see Figure S9c-d). Because of these reasons the comparability with the results shown in **Figure 2** is not given and annealing temperatures higher than 160 °C are not further discussed here.

We further investigated the anisotropic optical properties of the fully annealed films (**regime III**). **Figure 3a** shows the absorption spectra of a spin-coated and annealed thin film recorded at different angles of incidence (θ) for p- and s-polarized light. For increasing angle of incidence, the spectra for p- and s-polarized illumination differ significantly from each other. The H-band absorption at 2.54 eV increases for increasing angle of incidence only for p-polarized excitation, whereas it is virtually absent at normal incidence for p- and generally for s-polarized excitation. At normal incidence the spectra for p- and s-polarized illumination are identical, as expected. The H-band absorption is thus only excited by an electric field that has an out-of-plane component with respect to the substrate (see sketch in **Figure 3c**). These results imply the presence of uniaxial anisotropy and reveal a molecular orientation in out-of-plane direction (z-direction) extended over the entire thin film. We confirmed this finding by variable angle spectroscopic ellipsometry (VASE). Since standard spectroscopic ellipsometry in reflection is inherently limited in its sensitivity to the extraordinary components, we included transmission data (intensity as well as ellipsometry) and performed a multi-sample batch analysis.[43,44] From this we obtained a uniaxial anisotropic complex refractive index consisting of ordinary (in-plane) and extra-ordinary (out-of-plane) components. These complex refractive indices [*n* (refractive index) and *k* (absorption coefficient)] are shown in **Figure 3b**.

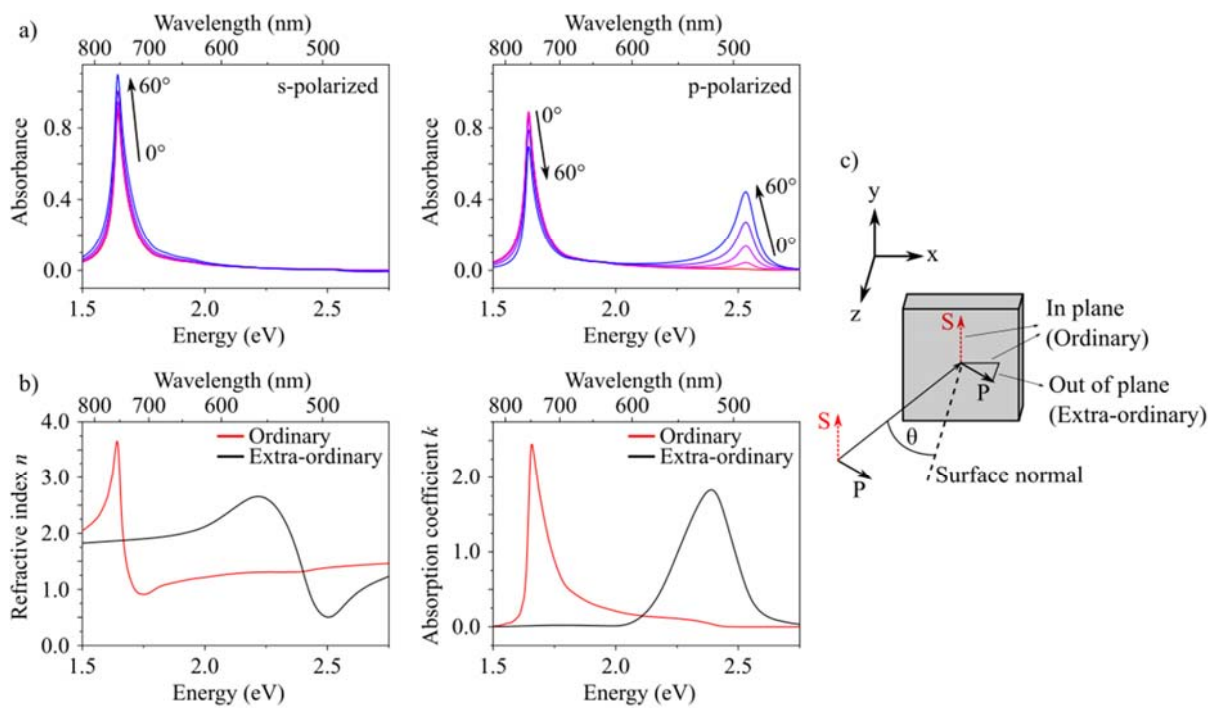

*Figure 3: Anisotropic optical absorption behavior of spin-coated and annealed (120 °C, 10 min.) HB238 thin films on glass substrates. (**a**) Absorption spectra at different angles of incidence (0° - 60° in 15° steps) for s- (left) and p-(right) polarized light. (**b**) Uniaxial anisotropic complex refractive index n (real part, refractive index, left) and k (imaginary part, absorption coefficient, right) determined by variable angle spectroscopic ellipsometry (VASE). (**c**) Scheme describing the electric field vectors of incident s- and p-polarized light relative to the substrate (xy-plane).*

The data in **Figure 3** shows that the J-TDM is oriented in the substrate plane (xy-plane). These findings match our expectation for aggregates with oblique angled monomer's TDMs in which the J- and H-band are polarized perpendicular to each other.[22,36,45] The lack of anisotropy in the substrate plane results from the random in-plane orientation of the crystallites (aggregates), since the macroscopic beam size in our experiments averages over many of them. The described morphology can also be observed in AFM images (see Figure S6f).

Note that the measured H-transition in the absorption spectra is blue-shifted by ca. 0.15 eV compared to the peak of the extraordinary absorption coefficient of the H-transition as determined by VASE. A similar phenomenon was found in a study of self-organized oligothiophene films where absorption peaks were calculated for transitions with different TDM orientations.[46] There, it was found that absorption peaks of transitions with out-of-plane TDMs are blue shifted compared to the peak in the complex refractive index used for the calculation. Absorption peaks of transitions with in-plane TDMs did not show this shift. We assume that these considerations apply to our findings as well.

A series of merocyanine molecules with identical π-systems, but different aliphatic side-chains, including HB238, was investigated recently by means of optical spectroscopy regarding their different aggregation behaviors.[18] There, only normal incidence measurements were performed, thus lacking the observation of HB238 H-transition absorption peaks. Instead, photoluminescence excitation measurements were performed, yielding information about the coupled nature of H- and J-transition for some of the investigated merocyanine aggregates.

Here, we prove the coupled nature of J- and H-bands by performing photoluminescence excitation measurements on the spin-coated and annealed films. A 0.8 NA objective was used for the measurement giving rise to AOIs of up to approximately 53°, making it possible to excite the H-transition. We found that the emission from the J-band was observed by exciting the H-band (see Figure S2), suggesting that our films consist of oblique-angled aggregates.

To gain further insights into the interdependence of structural and optical properties of HB238 spin-coated films, we computed via first-principles plane-wave based methods (see details in Experimental techniques and methods) the absorption spectrum of a HB238 single crystal. The crystals grown in water/acetone dispersion (**Figure 1**) were too small for the determination of the single crystal structure to be used in the first principle calculations. HB238 shows pronounced polymorphism, depending on the crystal growth conditions.[18,47] Single crystals grown from mesitylene solution[47] represent the properties of the aggregates investigated in this study. Therefore, we assume a similar crystal structure amongst the thin films and this single crystal (see Figure S3 and SI part I for details).

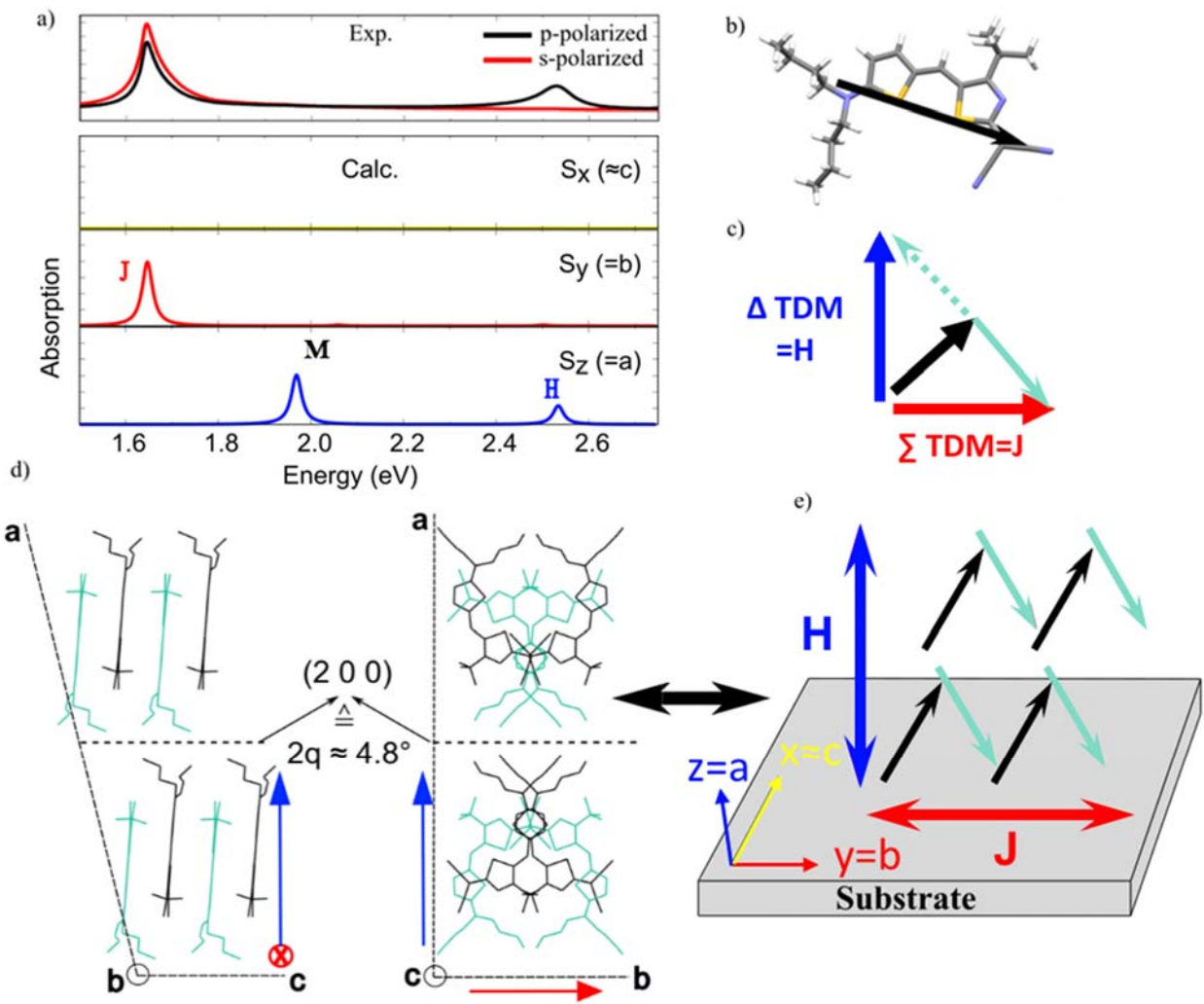

*Figure 4*: Calculated HB238 single crystal absorption spectrum in comparison with experiment. (*a*) Experimental absorption of a spin-coated and annealed (120 °C) thin film for p- (black) and s-polarized (red) light at an angle of incidence of 45° (top) and calculated polarized absorption spectra for HB238 single crystal (bottom). Yellow line – spectrum for light polarized along the Cartesian x-axis (i.e., similar to c crystallographic axis), red line – spectrum for light polarized along the Cartesian y-axis (i.e., b crystallographic axis), blue line – spectrum for light polarized along the Cartesian z-axis (i.e., a crystallographic axis). (*b*) HB238 molecular structure along with the calculated TDM (represented by black arrow). (*c*) Schematic representation of the perpendicular orientation of the TDM sum (red arrow) and difference (blue arrow) of two oblique angled TDMs (black and light blue). The concept is adopted from ref. 22. (*d*) Two representations of the single crystal unit cell showing the molecular packing. (*e*) Orientation of molecular TDMs (black and light blue arrows), and resulting J- and H response (red and blue double arrows) relative to the substrate. In first approximation the substrate plane (see text and **Figure 3**) closely coincides with the bc plane of the unit cell, leading to an overall out-of-plane (perpendicular) orientation for the H-band TDM and an in-plane orientation for the J-band TDM, i.e., edge-on orientation of the molecules.

**Figure 4a** displays the comparison between the experimental polarization-resolved absorption spectra of spin-coated and annealed (120 °C) thin films at 45° angle of incidence, and the calculated polarization-resolved absorption spectra of the single crystal obtained from mesitylene solution. The good agreement between the experimental and calculated spectra supports our assumption that the crystal structure of the single crystal grown from mesitylene solution is identical to the one in the thin films. By analyzing the induced charge-density,[48,49] namely the variation of the electron density upon an applied electric field, we can derive the spatial distribution of the exciton bands. The J-band at approximately 1.65 eV energy is related to a *delocalized* exciton whose induced charge density is polarized along the Cartesian y-axis (i.e., the *b* crystallographic axis). The identified H-band at approximately 2.5 eV energy, on the other hand, corresponds to a *delocalized* exciton along the Cartesian z-axis (i.e., the *a* crystallographic axis). The calculation shows an additional band around 1.9 eV for light polarized along the Cartesian z-axis. The induced charge density of this band corresponds to a *localized* exciton. This band is not observed in the thin film experiments. A reason for this could be related to the level of theory used here, which can lead to an overestimation of the intensity of some bands. Interestingly, this band can be observed in the absorption spectrum of dispersed aggregates (see **Figure 1b**). Due to its energetic position (compare with **Figure 1a**) and localized nature, we attribute this band to non-aggregated monomers.

Given the good agreement between experimental and calculated absorption spectra, we can discuss the molecular organization of HB238 molecules in the spin-coated and annealed thin films relative to the substrate surface. Our calculated results can be linked to the molecular orientation inside the single crystal structure by regarding the intuitive molecular exciton model of Kasha *et al.*[22]. The monoclinic HB238 single crystal structure shows a shifted, anti-parallel packing between merocyanines whose backbones, and thus TDMs (see **Figure 4b**) are oriented with an oblique angle relative to each other (black and red arrows in **Figure 4e**). For oblique angled TDMs, the TDMs for J- and H-transitions can be constructed by the vector sum and difference of the TDMs of the individual molecules, respectively, and are inevitably oriented perpendicular to each other, which is schematically depicted in **Figure 4c** for only two molecular TDMs for simplicity. Indeed, in our calculation, the J-band is polarized along the y Cartesian axis, coinciding with the *b* crystallographic axis, and the H-band is polarized along the z Cartesian axis, coinciding with the *a* crystallographic axis. We conclude the *bc* crystallographic plane is almost parallel to the substrate plane (xy). In this way the resulting orientation of the molecules relative to the substrate leads to a J- and H-transition oriented parallel, and perpendicular to the substrate plane, respectively, which agrees well with the measured ordinary and extraordinary complex refractive indices of the spin-coated and fully annealed films (see **Figure 3b**).

The supramolecular orientation presented in **Figure 4d** and **e** suggests high charge carrier mobilities in the plane of the substrate direction, due to the nearly perfect edge-on orientation of the molecules with respect to the substrate. In order to quantify the influence of aggregate formation on lateral charge carrier mobility, organic field effect transistors (OFETs) were fabricated by spin coating and annealing HB238 thin films (see details in Experimental techniques and methods). The gate dielectric is silicon-dioxide, i.e., chemically identical to the float glass and fused silica substrates used for the spectroscopic experiments shown in **Figure 2** and **Figure 3**. The aggregation behavior of HB238 on the OFET and float glass substrates was confirmed independently by 2θ XRD (Figure S3a and S4) and AFM measurements (Figures S5 – S9) and found to be essentially identical. Further, reflectance spectra of the OFET-samples were also measured for every annealing temperature at 50° AOI with p-polarized light, revealing a similar optically anisotropic behavior as for the studies on glass substrates (see Figure S18).

The samples used for the charge transport studies were annealed identically with the spectroscopy experiments on glass substrates. For every annealing temperature a separate sample was fabricated. The hole mobility was extracted from measurements of the drain-current in the linear regime for all samples and in the saturation regime for samples annealed at 120 °C or higher. The transistors did not show saturated drain currents for lower annealing temperatures (see Figures S19 and S20). The hole mobilities are shown in **Figure 5a** as a function of the annealing temperature. As in the spectroscopic investigations, the samples can be ordered in the same three regimes **I - III**. To visualize the correlation of aggregate formation and charge carrier transport, the hole mobility (**Figure 5a**) is shown as a function of $N_{coh}$ (**Figure 2b**) in **Figure 5b**.

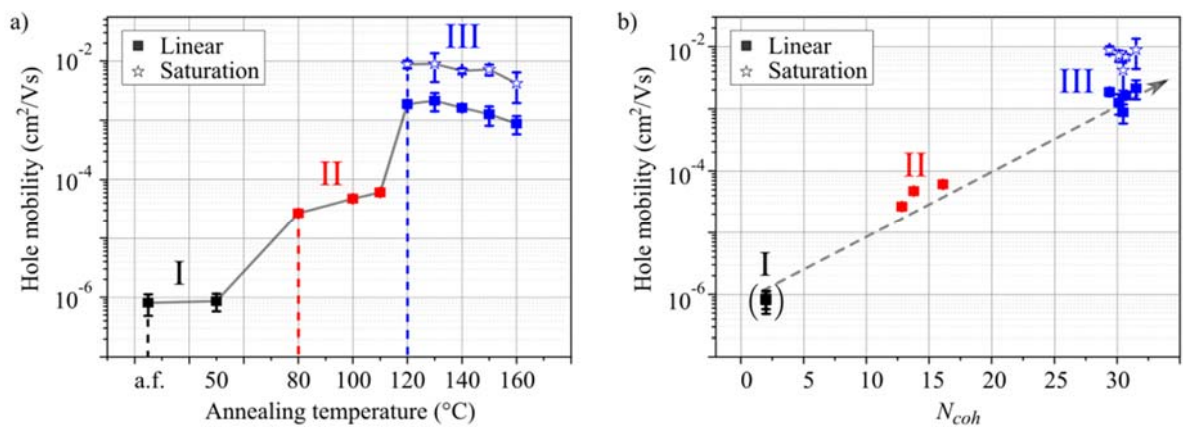

*Figure 5*: Hole mobility as a function of molecular aggregation. (**a**) Experimental hole mobilities of HB238 thin films spin-coated on OFET substrates for a non-annealed sample [as fabricated (a.f.)] and for samples after annealing at different temperatures between 50 °C and 160 °C for 10 minutes each. (**b**) Hole-mobilities as a function of the estimated number of coherently coupled molecules extracted from the absorption spectra in **Figure 2a**. The dashed grey arrow is a guide to the eye.

As shown in **Figure 5a,** the charge carrier mobility increases by more than three orders of magnitude as a consequence of the annealing-induced aggregation.[47] The unannealed sample and the sample annealed at 50 °C exhibit low mobilities of approximately $10^{-6}$ cm$^2$/Vs (**regime I**). Upon annealing at 80 °C, 100 °C, or 110 °C the mobility increases by ca. one order of magnitude (**regime II**). Finally, annealing the samples at 120 °C or higher leads to a drastic increase of charge carrier mobility by more than three orders of magnitude in total, up to approximately $10^{-3}$ cm$^2$/Vs in the linear and even $10^{-2}$ cm$^2$/Vs in the saturation regime and remains approximately constant for higher annealing temperatures (**regime III**).

As observed in **Figure 5b** the measured hole mobility and $N_{coh}$ appear to be strongly correlated. Both increase significantly during the transitions between the regimes, only gradually within **regime II**, and remain practically constant in **regime III**. The increasing hole mobility with increasing $N_{coh}$ can be attributed to a higher degree of crystallization with higher annealing temperature, as also evident from the XRD data (Figure S4). In crystalline domains charge transport proceeds via band transport, as opposed to the much slower hopping transport in amorphous domains. We attribute the saturation of both, $N_{coh}$ and hole mobility in **regime III** to a constant defect density for spin-coated thin films of HB238 after the threshold annealing temperature is reached (transition from **regime II** to **III**). This explanation is rationalized, by the fact that defects represent a localization site for excitons. At the same time, defects lead to either trap states or potential wells at which charge transport is hindered and might proceed by hopping.

The combined spectroscopic, XRD, and charge transport data proves that in **regime III** the spin-coated HB238 thin films entirely consist of nearly perfect edge-on oriented molecules (out-of-plane anisotropy), which significantly increases charge carrier mobility in a FET geometry. However, the crystallite orientation within the substrate plane is random (in-plane isotropy; see Figures S6-S8). In a computational study on the exactly same crystal structure, it was found, that the most efficient transport pathway for holes is along the *b* crystallographic axis,[47] which interestingly coincides with the orientation of the TDM of the J-transition, schematically shown in **Figure 4**. This coincidence allows the determination of the direction of most efficient hole transport by polarization-resolved microspectroscopic methods.

By growing larger and in-plane oriented crystalline domains the hole mobility can potentially be enhanced. The morphology of thin films can be improved by utilization of ordered substrates, acting as a "template" for the organic thin films. In order to obtain larger ordered domains, HOPG is used as a highly ordered substrate. We fabricated HB238 thin films on HOPG by thermal vapor deposition. The morphology and optical properties of these films were analyzed by means of spatially correlated AFM and PL microspectroscopy. AFM- and PL-micrographs were recorded at the same area of the sample. Furthermore, PL-spectra were measured at different positions within these areas. **Figure 6a** and **c** show AFM topography images of a mono- and a multilayer HB238 thin film on HOPG, respectively (see Experimental techniques and methods for details). Corresponding linear height profiles, indicated by a white bar, are plotted in panel **e**. The positions where linear height profiles were extracted are highlighted by a white rectangle in panel **a** and **c**, displayed enlarged in panel **e**. The height profiles display approximately 2 nm steps for both the mono- and multilayer. Comparing this with the dimensions of the HB238 crystal structure discussed above (see panel **e**), we conclude that also here the molecules are oriented edge-on as in the spin-coated thin films in **regime III**. The step height is about half of the height of the unit cell if the crystallographic *bc* plane is parallel to the substrate. This corresponds to one layer of edge-on oriented molecules. The multilayer has a "wedding cake" like structure with stacked layers of equal height and a from layer-to-layer decreasing area as elucidated in **Figure 6e**. Additionally, thicker crystallites are found on the surface of mono- and multilayer samples. The surface step heights as well as the spectroscopic properties of those differ from the dominating edge-on structure. Here the height profiles show approximately 0.5 nm steps. We therefore conclude that the molecules in these crystallites must be arranged in a "face-on" stacking (lying flat on the surface, see Figure S23).

The evaporated films on HOPG and the spin-coated and annealed thin films on glass both with edge-on domains show similar optical properties. Polarization-resolved reflection spectra for different AOI (see Figure S22) display similar uniaxial anisotropy. Furthermore, 2θ XRD measurements confirm the same crystal phase in dispersed aggregates, spin-coated and annealed films, and on HOPG (see Figure S3a). Reflection and XRD measurements were performed over a macroscopic spot size and therefore include the face-on crystallites, too. However, since the HOPG surface is covered dominantly by the edge-on structures (as shown in **Figure 6a** and **c**), we conclude that the reflection spectra and XRD-scans display mainly the properties of the edge-on structure. Finally, the PL spectra of the evaporated thin films on HOPG and of the spin-coated and annealed thin films on $SiO_2$ are very similar and show only the characteristic J-transition emission at approximately 1.65 eV as shown in **Figure 6f**. These results prove that the same aggregation type is present in all samples.

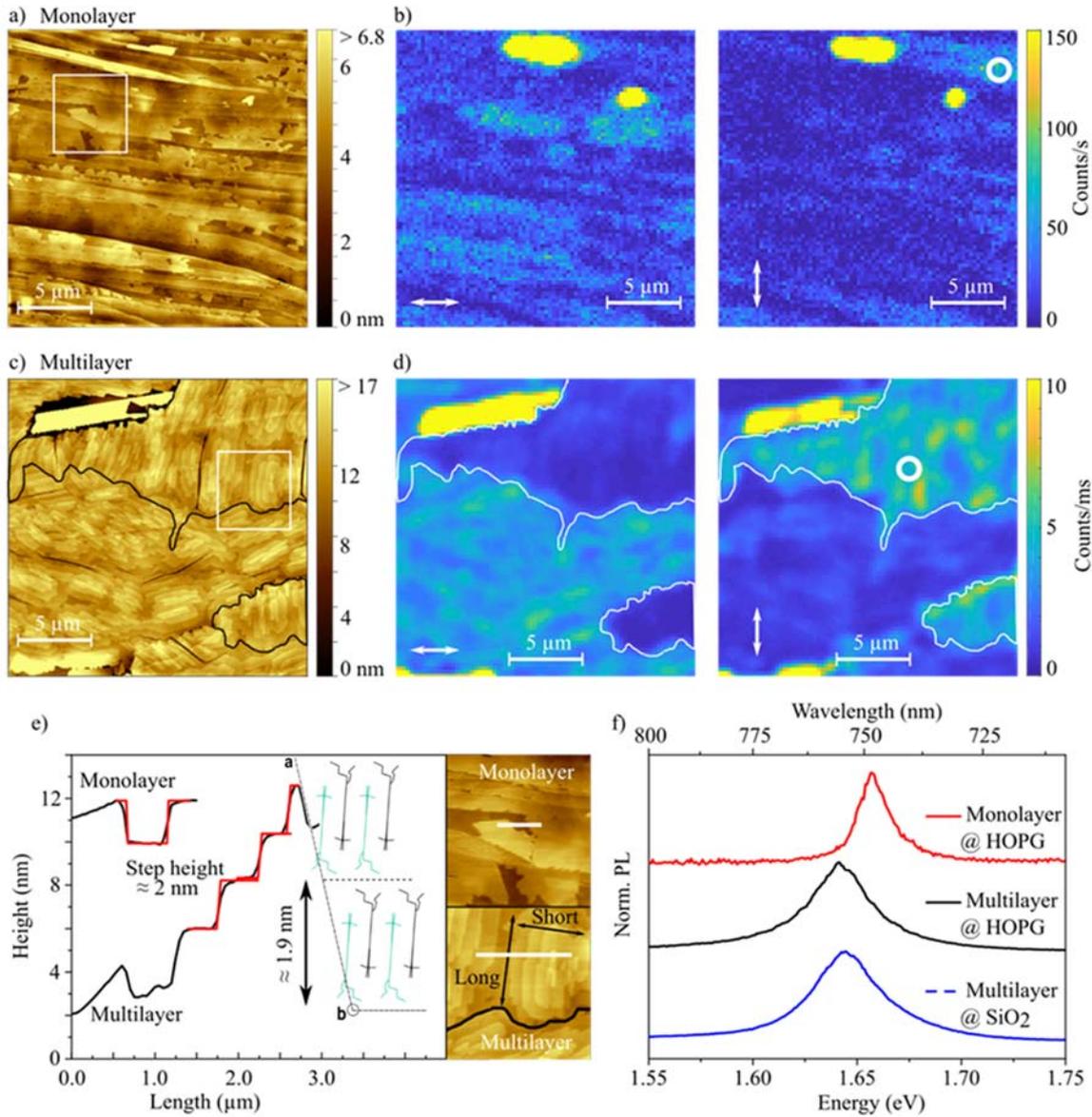

*Figure 6*: Correlated AFM and polarization-resolved PL study. (*a*) and (*c*) AFM images of a thermally evaporated HB238 mono- and multilayer on HOPG. The black solid lines in panel (*c*) indicate the domain boundaries and are guides to the eye. (*b*) and (*d*) Corresponding PL micrographs at the same position for different linear polarizations of the excitation light (indicated by white arrows). The white solid lines in panels (*d*) indicate the same domain boundaries as the black solid lines in panel (*c*). (*e*) Linear height profiles measured at the positions of the white bars indicated in the close-ups of the AFM height images on the right of panel (*e*). The areas of the close-ups are indicated with white squares in panels (*a*) and (*c*). The molecular orientation relative to the HOPG substrate surface is shown in the inset. (*f*) PL spectra of HB238 mono- (red) and multilayer (black) on HOPG, taken at positions indicated with white circles in panel *b* and *d*, compared to PL spectra of a spin-coated film on an OFET substrate annealed at 120 °C (blue).

**Figure 6b** and **d** show PL-micrographs for different linear polarizations of the excitation light indicated with white arrows, for the mono- and multilayer sample respectively. White circles indicate positions at which spectra were recorded. The spectra are shown as solid lines in panel **f**. The observed micrographs are remarkably different for different excitation polarizations, best seen for the multilayer in panel **d**. Since the *a* crystallographic axis is oriented almost perpendicular to the substrate surface throughout the entire film (apart from face-on crystallites) and the J-transition polarization axis coincides with the *b* crystallographic axis (see **Figure 4c**) we can determine the crystal unit cell orientation on the substrate surface from the PL brightness relative to the excitation polarization. For example, the comparably large edge-on domain on the top right in **Figure 6d** is oriented with the b crystallographic axis roughly vertical with respect to the image, indicated by a much brighter PL for vertical excitation polarization. Additionally, by comparing with the AFM image, a similarly vertical crystallite in-plane orientation is identified, determined by the relative orientation of their long-edges, shown in panel **c** and in the inset in panel **e**. The black solid lines in the multilayer AFM image (panel **c**) separate the observed domains and are guides to the eye. With this information it is further possible to determine the directions or paths of most efficient hole mobility, since these coincide as well with the b crystallographic axis, just by observing the AFM or PL images.

Our combined investigations reveal that the aggregation state and molecular orientation determined for the spin-coated and annealed films, which improved charge transport properties by three orders of magnitude, is maintained in the ordered films on HOPG. The grain size of the oriented domains in the thin films on HOPG is orders of magnitude larger (10s - 100s of $\mu m^2$, see **Figure 6a** and **c**) than the "crystallites" in spin-coated and annealed films on OFET substrates (< 1 $\mu m^2$, suggested by AFM topography in Figures S6e – S8a), which most likely results in significantly increased charge carrier mobility in thin films on HOPG. Since the lateral grain size is increased from less than 1 µm in the spin-coated films on amorphous substrates to more than 10 µm in the evaporated films on the templating substrate HOPG itself, we assume a likewise enhancement of lateral charge carrier mobility by more than an order of magnitude. However, due to the electric conductivity of HOPG it was not possible to fabricate OFETs from these films. The low emission intensity of the monolayer makes it more difficult to observe this effect in the polarized PL-scans (**Figure 6b**). In addition, the physical orientation (i.e. the orientation of the long edge) cannot be determined for the monolayer due to the missing additional layers. However, we find large contiguous structures of at least 10s of µm$^2$ in size, limited by the substrate's step edges (**Figure 6a**).

As a comparison an HB238 thin film multilayer was evaporated on float glass using the exact same conditions as for the multilayer on HOPG. We observe the same aggregation state as on HOPG. However the domains are much smaller (around 5 µm$^2$), randomly oriented and separated by pronounced grain boundaries (see Figure S24 for details). This proves, that the large, directionally oriented domains found on HOPG are indeed a result of the ordered HOPG surface.

**Figure 6f** shows PL-spectra from the evaporated mono- and multilayer thin film on HOPG, compared with PL spectra from a spin-coated and annealed (120 °C, 10 min.) film on SiO$_2$ (OFET substrate). The energetic positions of the J-band emission are similar between all samples and the slight energetic shifts are most likely attributed to dielectric phenomena, driven by the different energetic surroundings of the associated HB238 molecules. The molecules in the multilayer films on HOPG and SiO$_2$ are mainly surrounded by other HB238 molecules and thus possess a similar transition energy. The molecules in the monolayer on HOPG are comparatively stronger interacting with the HOPG surface and we observe slightly blue-shifted emission. It can be recognized that the peak-width of the HOPG-monolayer PL-spectrum is significantly smaller (FWHM = 18 meV (145 cm$^{-1}$)) than for the other spectra (FWHM = 35 − 40 meV (282 − 323 cm$^{-1}$)). We attribute this finding to less disorder in the monolayer grown on HOPG. It is known from theoretical considerations, that in molecular aggregates disorder induces mixing of the bright lowest state with higher lying states, as well as localization of excitons, both broadening the electronic transition.[50–52] It can be assumed that the monolayer on HOPG exhibits higher order and thus a higher degree of exciton delocalization and potentially charge carrier mobility than the other samples.

## Conclusion & Outlook

We have studied the aggregation of the merocyanine HB238 and its impact on charge carrier mobility. We find that the mobility in an OFET with a spin-coated thin film of the organic semiconductor can be significantly increased by thermal annealing and accompanied aggregation.

Beneficial for charge carrier transport, these films adopt a preferential out-of-plane molecular orientation and thus co-facial orientation of the molecular π-systems. To further enhance lateral charge transport properties, the in-plane morphological order of HB238 thin films was significantly improved by the utilization of HOPG as substrate. Evaporation of HB238 on HOPG led to ordered growth and we observe out-of-plane, as well as in-plane orientation in aggregated domains of up to 100s of μm$^2$ in size. Especially the monolayer edge-on structure on HOPG displays a high degree of order as evident from very sharp PL-spectra. We expect that the lateral charge carrier mobility of HB238 thin films on ordered HOPG substrates is enhanced by more than an order of magnitude compared to the films on amorphous substrates, according to the likewise enlargement of lateral grain size. This is a promising result on the way to high charge carrier mobility in ordered organic aggregate thin films.

Our results show that the molecular arrangement in organic semiconductor thin films significantly influences device performance. The control over the molecular arrangement and grain size is one of the most important goals in the development of efficient organic electronic devices, which, as our results show, can be achieved by choosing a suitable templating substrate. The significantly increased domain size and order, as evidenced by the spectrally sharp PL-spectra, suggest that suitable templates may be used to realize organic electronics devices with superior performance. The here studied thin films further have interesting optical properties. For example, the highly anisotropic nature of HB238 thin films was recently applied in polarization-dependent strong light-matter coupling.[53] The anisotropy may find further applications in polarization-sensitive devices and structures.

## Experimental Techniques and Methods

**Sample Fabrication**

*Solution and Dispersion*

The merocyanine HB238 was synthesized following ref. [6]. It was dissolved in acetone (*Fisher Chemicals*, HPLC grade) with a concentration of c = 1 · 10$^{-5}$ mol/l resulting in HB238 being present in the monomeric form in solution (see Figure S1). Aggregates in dispersion were prepared by transferring 0.1 ml of this solution into 0.9 ml of deionized water. We observed an instantaneous color change from deep blue to light purple.

The sample of dispersed aggregates for 2θ XRD scans (Figure S3a) was produced in the same way by using an HB238/acetone solution with a concentration of c = 1 · $10^{-2}$ mol/l. In this way HB238 aggregates of purple color precipitate visibly by eye. The precipitate was placed on a float glass substrate with a pipette and allowed to dry overnight.

*Spin-coated films on silicon dioxide surfaces*

All spin-coated samples in this work were fabricated from HB238 in chloroform (*Fisher Chemicals*, HPLC grade) solutions with a concentration of c = 1 · $10^{-2}$ mol/l. Spin coating was then performed by dropping 0.1 ml of this solution onto the substrates (float glass, *Paul Marienfeld GmbH & Co. KG*) for transmission spectroscopy and luminescence excitation, ellipsometry (quartz glass, *Nano Quarz Wafer GmbH*), and transistor-measurements and PL-spectroscopy (OFET-substrates, *Fraunhofer IPMS, Dresden*). Subsequently the spin coater was started (static dispense). All samples were spun at 3000 rpm speed with 3000 rpm/s acceleration, and for 60 seconds. With these parameters film thicknesses of approximately 20 nm were obtained as determined using atomic force microscopy (AFM). Subsequent annealing was performed by placing the substrates on a preheated hot plate for 10 minutes directly after the spin coating process. The used temperatures are given in the main text. The OFET substrates were ozonized for ten minutes before use to enhance the surface polarity and thus to improve HB238 film formation. For every annealing temperature a separate sample was prepared.

*Evaporated thin films on HOPG and float glass*

The HOPG substrates were cleaved with scotch tape right before use in nitrogen atmosphere. Evaporation of HB238 was performed in a custom-built *CreaPhys GmbH* vacuum system with a constant evaporation rate of 0.03 Å/s at a pressure of 2 · $10^{-7}$ mbar onto HOPG substrates heated to 80 °C. For the mono- and multilayer samples, a nominal film thickness of 2.5 nm and 20 nm was evaporated, respectively. For the evaporation on float glass a nominal film thickness of 20 nm was deposited under the same conditions described above, using the same float glass substrates as in the spin coating study.

**Optical Spectroscopy**

Transmission and reflection spectroscopy were performed with a *PerkinElmer Lambda 1050* Spectrophotometer equipped with a removable linear polarizer and a rotatable sample mount, using a transmission 3-detector module or a universal reflectance accessory module. Both transmission and reflection spectroscopy were performed with macroscopic beam sizes of approximately 10 mm$^2$ area. All absorption spectra shown in this work were corrected by dividing the measured intensities of the fabricated samples by the blank substrate/cuvette spectra under the same experimental conditions (i.e. angle of incidence and polarization or pure solvent).

H- and J-transition peaks are known to be asymmetric. This is because H- (J-) transitions are energetically located at the exciton band-top (-bottom), thus only mixing with energetically lower (higher) lying states, which broadens the observed spectral peaks asymmetrically.[28] Thus, absorption peak parameters were determined by fitting the data using asymmetric pseudo-Voigt profiles, having different widths on the high and low energy wing of the peak. The absolute peak-width is then calculated as the mean of both widths. The data to be fitted was chosen such that for the J-transition, the absorption up to the photon energy corresponding to the half of the peak maximum at the high energy side of the peak was considered. Similarly, for the H-transition, the absorption from the photon energy corresponding to the half of the peak maximum at the low energy side of the peak was considered. This minimizes the influence of spectrally overlapping transitions (see SI for exceptions).

The PL microspectroscopy analysis was performed with an in-house built confocal microscopy setup, equipped with a 100x magnification, 0.8 NA objective (*Olympus*). The sample is positioned and scanned using a three-dimensional nanopositioning stage. The collected photoluminescence light is detected using a single-photon counting detector (*MPD*) for PL-imaging or a spectrometer (*Princeton Instruments, IsoPlane SCT 320*) for PL-spectroscopy. The excitation source is a tunable white light source (*NKT Photonics, SuperK EXTREME*) equipped with a spectral filter system (*NKT Photonics, SuperK VARIA*). The excitation light is focused onto the sample using the microscope objective. The PL spectra were recorded with an excitation wavelength of 490 nm (the emission was filtered using a 635 nm long pass filter) and 555 nm (568 nm long pass filter) for the aggregate and monomer spectra shown in **Figure 1a** and **b**, respectively. All thin film PL-spectra in this work were recorded with an excitation wavelength of 600 nm (635 nm long pass filter). All spectra on thin films were recorded utilizing an excitation power below 1.5 µW to minimize photo-bleaching. Fitting of J-transition emission peaks was performed applying the same procedure as for absorption peaks mentioned before, where here the fitting boundaries were chosen at the energy value which corresponds to half the peak maximum at the low energy side.

The luminescence excitation analysis (Figure S2) was executed using the same PL-setup. Spectra were collected on the same sample position for different excitation wavelengths (460 nm – 730 nm, in 10 nm steps), and a 750 nm long pass emission filter was used in the emission path. To determine the total emitted signal for each excitation wavelength, the emission spectra were fitted using a Lorentz peak function for the J-transition emission peak. The total emitted power was corrected with the power of the excitation beam. The excitation power was kept below 0.15 μW for all spectra, to minimize photo-bleaching. At the end of the experiment the photo-bleaching was approximately 25 %. Photo-bleaching can thus not be neglected but the general shape of the luminescence excitation spectrum is revealed.

The complex refractive index of the films were determined using variable angle spectroscopic ellipsometry using an M-2000-XI (J.A. Woollam) and analyzed using the CompleteEASE software (version 6). Standard spectroscopic ellipsometry in reflection (SEr), normal incidence transmission intensity as well as variable angle transmission spectroscopic ellipsometry (SEt) scans have been simultaneously fitted using the multi-sample analysis option. The angles of incidence for SEr were from 45° to 75° in steps of 5°, and from 0° to 65° in steps of 5° for SEt. The spectral range was from 245 nm 1690 nm. The transparent regime from 900 nm to 1690 nm was used to determine the sample layer thickness by a Cauchy model fit also accounting for incoherent backside reflections. Conversion to uniaxial anisotropic B-spline and wavelength expansion fit was performed subsequently with fixed layer thickness. Node spacing was set to 0.1 eV (0.2 eV) for the in-plane (out-of-plane) components and further lowered in the absorbing spectral regions. An MSE of 6.2 was obtained for the multi-sample analysis. The model-free B-spline fit gives a very good match of the transmission intensity data.

**Atomic Force Microscopy (AFM)**

All measurements were performed with an *Asylum* research AFM (*Oxford Instruments*, *MFP−3D Infinity*) in alternating contact mode utilizing AC200TS cantilevers (*Olympus*). Data evaluation was performed using the software *Gwyddion*. Due to the undulating nature of the HOPG surface, the associated data images were processed by subtracting a polynomial background to improve the visual contrast resulting in the images shown in **Figure 6a** and **c**. The evaluation of height profiles was performed before background subtraction.

**Electrical characterization**

Transistor measurements were performed using a *Keithley, 4200A-SCS Parameter Analyzer*. The OFET-substrates have 16 drain-source electrode pairs. The substrate is highly doped silicon with a 300nm thick gate oxide. The devices have channel-lengths from 2.5 µm to 20 µm and a channel−width of 1 cm. In this study, for every sample at least three devices with a channel-length of 20 µm were used for evaluation. The charge carrier mobility was determined from the transfer characteristics using the formulas[29]

$$\mu_{linear} = \frac{\partial I_D}{\partial V_G} \frac{L}{W C_{SiO_2} V_D}, \text{ and} \qquad \text{(eq. 1)}$$

$$\mu_{saturation} = \left(\frac{\partial \sqrt{I_D}}{\partial V_G}\right)^2 \frac{2L}{W C_{SiO_2}}, \qquad \text{(eq. 2)}$$

where $\mu_{linear}$ and $\mu_{saturation}$ are the hole mobilities in the linear and saturation regime, respectively, $I_D$ is the drain-current, $V_G$ is the gate-voltage, $L$ and $W$ are the channel-length and – width, $C_{SiO_2}$ is the gate-dielectric capacitance per unit area, and $V_D$ is the drain-voltage. The derivatives in **eqs. 1** and **2** were evaluated from a linear fit to the measured dependence of the drain-current or the square root of the drain current on the gate-voltage in the linear regime (fitting range of gate voltage from −20 V to −50 V), or saturation regime (fitting range of gate voltage from −20 V to −40 V), respectively (see Figures S19 and S20). The corresponding transfer characteristics were recorded from +10 V to -50 V gate voltage at a constant drain voltage of -10 V or -130 V in the linear and saturation regime, respectively.

**X-ray Diffraction (XRD)**

The diffraction patterns were recorded with an *Empyrean* diffractometer (*Malvern Panalytical*) in 2θ geometry. The measurements were performed with a Cu K-α X-Ray source using a fixed divergence slit with a slit size of 0.0286. The structure of single crystals was determined using a *Bruker D8-Venture* with kappa-geometry equipped with a Cu *Microfocus* X-Ray source and a *PhotonIII M14* detector. The disordered solvent molecules included in the crystal were treated with the software *PLATON SQUEEZE*[30], and single crystal structure determination and refinement were performed using the software *SHELXT*[31] and *SHELXL*[32], respectively.

**Quantum chemical calculations**

Density functional theory (DFT) calculations by using the range separated functional ωB97X-D and 6-311G** basis set were employed to obtain the equilibrium molecular structure and vibrational force field of HB238. Excited states, namely vertical transitions, were computed by TD-DFT calculations. The dipole active excited state S1 was optimized, and the vibrational force field was computed at the same level of theory. In such a way, the Franck-Condon (FC) factors were computed considering the ground and excited state equilibrium geometries and force fields. The computed vibronic absorption spectrum in gas-phase is reported in Figure S1. DFT calculations were performed using Gaussian16 (B.01 version).[33]

The dynamic polarizabilities, the absorption spectrum and the induced charge densities of HB238 single crystal were computed via pseudopotential plane-wave DFT calculations using the suite of codes QuantumESPRESSO[34] coupled with the turboTDDFT [35]. The latter efficiently implements a Liouville−Lanczos approach to time-dependent density functional theory for the calculation of the absorption spectra in the frequency domain. For such a calculation, the PBE-generalized gradient exchange−correlation functional and ultrasoft pseudopotentials from the PS library were used. Single-particle wave functions (charge density) are expanded in plane waves up to an energy cutoff of 70 Ry (700 Ry). Only Γ point has been considered for Brillouin zone sampling in the reciprocal space. Dynamic polarizabilities were computed with respect to the Cartesian coordinates, rather than crystallographic ones. Therefore, the z-axis coincides with **a**, y-axis with **b**, while x-axis differs from **c**, the latter showing an angle with respect to the **ab** plane (monoclinic unit cell, see **Figure 4d**).

# Notes

The authors declare no competing financial interest.

# Acknowledgments

This work was funded by the RTG-2591 "TIDE ─ Template-designed Organic Electronics" (Deutsche Forschungsgemeinschaft, DFG) and by the University of Cologne through the Institutional Strategy of the University of Cologne within the German Excellence Initiative (QM2). We thank F. C. Spano (Temple University) for very fruitful discussions about the estimation of the coherence length and S. Olthof (University of Cologne) for support discussing the XRD data. We acknowledge useful discussions with M. Reimer (University of Cologne).the coherence length and S. Olthof (University of Cologne) for support discussing the XRD data. We acknowledge useful discussions with M. Reimer (University of Cologne).